\documentclass{iaus}
\usepackage{graphicx}

\newcommand{\Msun}{\mathrm{M}_{\odot}}
\newcommand{\feh}{\mathrm{[Fe/H]}}
\def\apj{\textit{ApJ}}
\def\mnras{\textit{MNRAS}}

\title[IAUS270.~~Formation of Globular Clusters]{
  Modeling Formation of Globular Clusters: Beacons of Galactic Star Formation}

\author[Oleg Y. Gnedin]{Oleg Y. Gnedin}

\affiliation{University of Michigan, Department of Astronomy,
    Ann Arbor, MI 48109, USA \\ {\tt ognedin@umich.edu}}

\pubyear{2010}
\volume{270}
\pagerange{1--4}
\jname{Computational Star Formation}
\editors{Alves, Elmegreen, Girart, Trimble, eds.}
\begin{document}

\maketitle

\begin{abstract}
Modern hydrodynamic simulations of galaxy formation are able to
predict accurately the rates and locations of the assembly of giant
molecular clouds in early galaxies.  These clouds could host star
clusters with the masses and sizes of real globular clusters.  I
describe current state-of-the-art simulations aimed at understanding
the origin of the cluster mass function and metallicity distribution.
Metallicity bimodality of globular cluster systems appears to be a
natural outcome of hierarchical formation and gradually declining
fraction of cold gas in galaxies.  Globular cluster formation was most
prominent at redshifts $z>3$, when massive star clusters may have
contributed as much as 20\% of all galactic star formation.
\end{abstract}

\firstsection
\section{Clues from Old and Young Star Clusters}

A self-consistent description of the formation of globular clusters
remains a challenge to theorists.  Most of the progress is driven by
observational discoveries.  The Hubble Space Telescope observations
have convincingly demonstrated one of the likely routes for the
formation of massive star clusters today -- in the mergers of gas-rich
galaxies.  These observations have also shown the differences between
the mass function of young clusters (power-law $dN/dM \propto M^{-2}$)
and old clusters (log-normal or broken power-law).

Surveys of the globular cluster systems of galaxies in the Virgo and
Fornax galaxy clusters have solidified the evidence for bimodal, and
even multimodal, color distribution in galaxies ranging from dwarf
disks to giant ellipticals (\cite{peng_etal08}).  This color
bimodality likely translates into a bimodal distribution of the
abundances of heavy elements such as iron.  We know this to be the
case in the Galaxy as well as in M31, where accurate spectral
measurements exist for a large fraction of the clusters.  The two most
frequently encountered modes are commonly called {\it blue}
(metal-poor) and {\it red} (metal-rich).

Detailed spectroscopy reveals a significant spread of ages of the red
clusters in the Galaxy, up to 6 Gyr (\cite{dotter_etal10}).  The
spread increases with metallicity and distance from the center.  The
age spread of the blue clusters is smaller, in the range 1-2 Gyr, and
is consistent with the measurement errors.

\section{Modeling the Formation of Globular Clusters is Hard}

The first attempt to model the formation of globular clusters within
the framework of hierarchical galaxy formation was by
\cite{beasley_etal02}.  Their semi-analytical model could reproduce
the metallicity bimodality only by assuming two separate prescriptions
for the blue and red clusters: blue clusters formed in quiescent disks
with an efficiency of 0.002 relative to field stars, whereas red
clusters formed in gas-rich mergers with a higher efficiency of 0.007.
The formation of blue clusters also had to be artificially halted
after $z=5$, so as not to dilute the bimodality.

\cite{moore_etal06} considered an idealized scenario for the formation
of blue globular clusters at high redshift, inside dark matter halos
that would eventually merge into the Galaxy, one cluster per halo.
They used the observed spatial distribution of the Galactic clusters
to constrain the formation epoch and found that the clusters would
need to form by $z \sim 12$, in relatively small halos.  Such an early
formation is inconsistent with the simultaneous requirements of high
mass and density for the parent molecular clouds to produce such dense
($\rho_* > 10^4\, \Msun$ pc$^{-3}$) and massive ($M > 10^5\, \Msun$)
clusters as observed.  This scenario also places stringent constraints
on the age spread of blue clusters to be less than 0.5 Gyr, which may
already be inconsistent with the available age measurements.  The
tension with observations of this scenario, and of its several
variants in the literature, probably indicates that globular clusters
cannot be simply associated with early dark matter halos and, instead,
must be studied as an integral part of galactic star formation.

\section{Globular Clusters Could Form in Protogalactic Disks}

\cite{kravtsov_gnedin05} used a hydrodynamic simulation of a Galactic
environment at redshifts $z > 3$ and found dense, massive gas clouds
within the protogalactic clumps.  These clouds assemble within the
self-gravitating disk of progenitor galaxies after gas-rich mergers.
The disk develops strong spiral arms, which further fragment into
separate molecular clouds located along the arms as beads on a string.
A working assumption, that the central high-density region of these
clouds formed a star cluster, results in the distributions of cluster
mass, size, and metallicity that are consistent with those of the
Galactic metal-poor clusters.  The high stellar density of Galactic
clusters restricts their parent clouds to be in relatively massive
progenitors, with the total mass $M_h > 10^9\ \Msun$.  The mass of the
molecular clouds increases with cosmic time, but the rate of mergers
declines steadily.  Therefore, the cluster formation efficiency peaks
during an extended epoch, $5 < z < 3$, when the Universe is still less
than 2 Gyr old.  The parent molecular clouds are massive enough to be
self-shielded from UV radiation, so that globular cluster formation
should be unaffected by the reionization of cosmic hydrogen at $z >
6$.  The mass function of model clusters is consistent with a power
law $dN/dM \propto M^{-2}$, similar to the local young star clusters.
The total mass of clusters formed in each progenitor is roughly
proportional to the available gas supply and the total mass, $M_{GC}
\sim 10^{-4}\ M_h$.

\cite{prieto_gnedin08} showed that subsequent mergers of the
progenitor galaxies would ensure the present distribution of the
globular cluster system is spheroidal, as observed, even though
initially all clusters form on nearly circular orbits.  Depending on
the subsequent trajectories of their host galaxies, clusters form
three main subsystems at present time.  {\it Disk clusters} formed in
the most massive progenitor that eventually hosts the present Galactic
disk.  These clusters are scattered into eccentric orbits by
perturbations from accreted galactic satellites.  {\it Inner halo
clusters} came from the now-disrupted satellite galaxies.  Their
orbits are inclined with respect to the Galactic disk and are fairly
isotropic.  {\it Outer halo clusters} are either still associated with
the surviving satellite galaxies, or were scattered away from their
hosts during close encounters with other satellites and consequently
appear isolated.

Following the scenario outlined above, \cite{muratov_gnedin10}
developed a semi-analytical model that aims to reproduce statistically
the metallicity distribution of the Galactic globular clusters.  The
formation of clusters is triggered during a merger of gas-rich
protogalaxies with the mass ratio 1:5 or higher, and during very early
mergers with any mass ratio when the cold gas fraction in the
progenitors is close to 100\%.  Model clusters are assigned the mean
metallicity of their host galaxies, which is calculated using the
observed galaxy stellar mass-metallicity relation.

\begin{figure}[t]
\centering
 \includegraphics[width=2.63in]{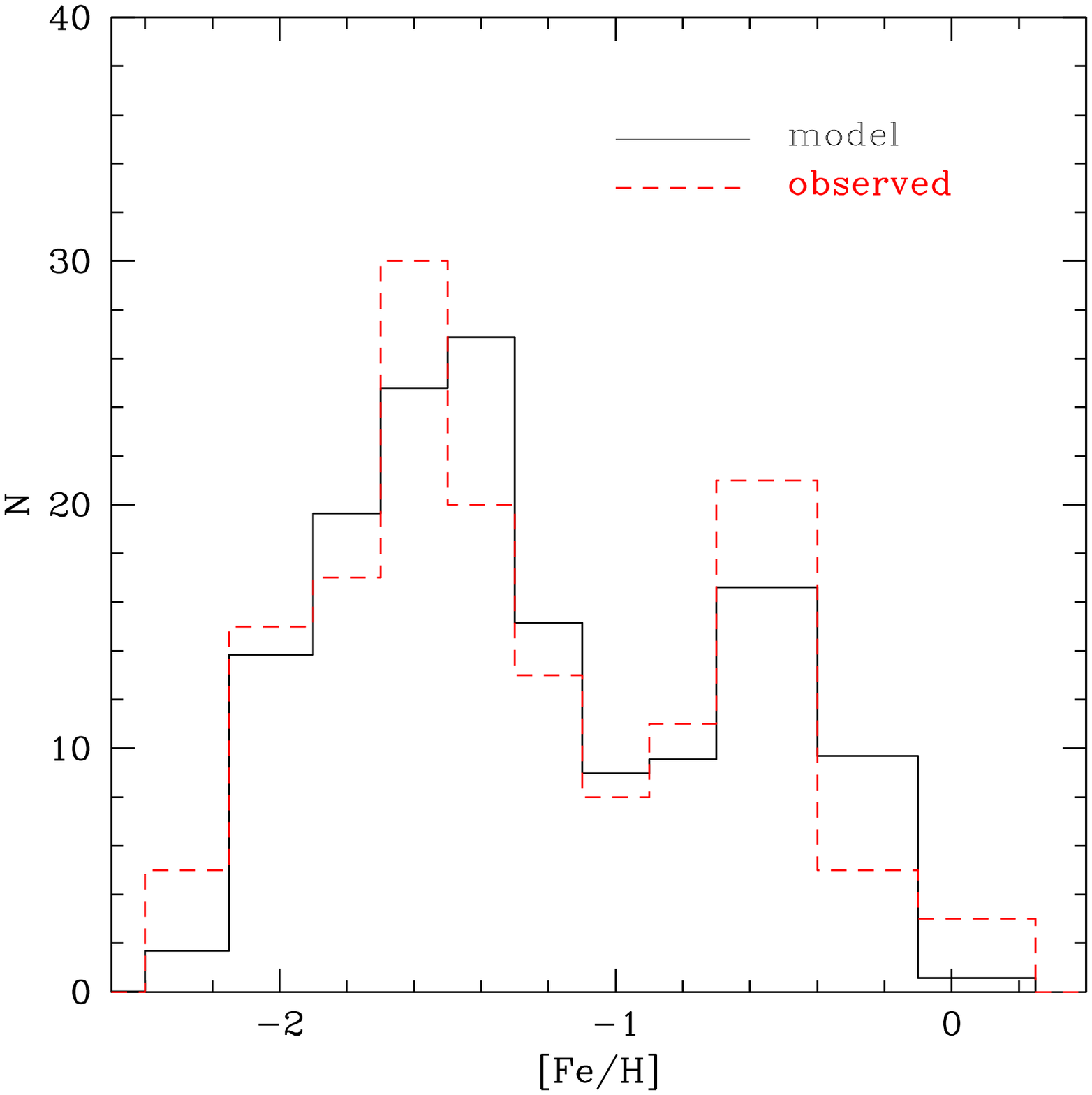}
 \includegraphics[width=2.63in]{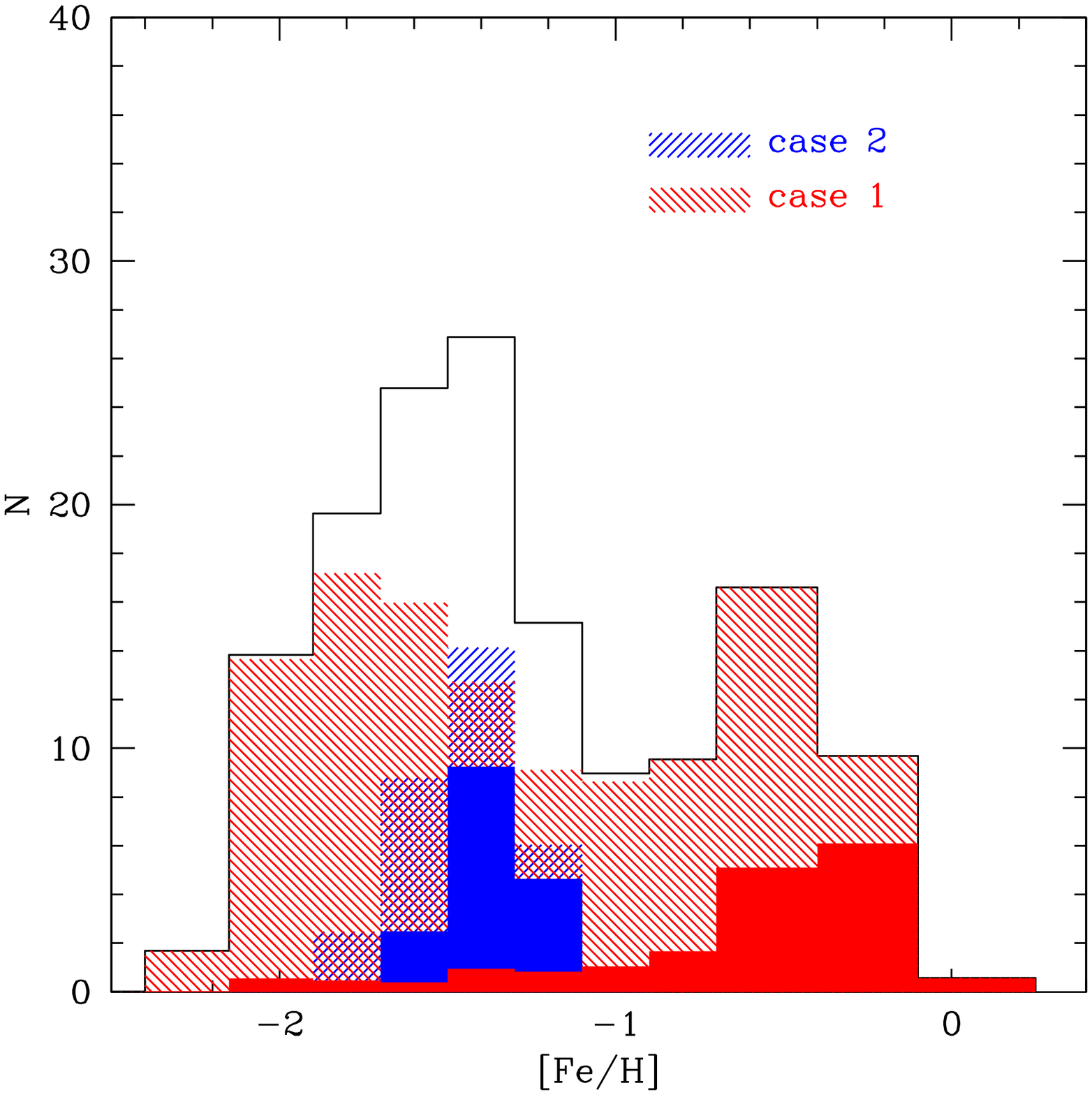}
\vspace{-0.4cm}
\caption{\small {\it Left:} Metallicities of model clusters that
  survived dynamical disruption until $z=0$, compared to the observed
  distribution of Galactic globular clusters.  {\it Right:} Model
  metallicity distribution split by the formation criterion: major
  mergers ({\tt case-1}) and early mergers ({\tt case-2}).  Filled
  histograms show clusters formed in the main Galactic disk.  From
  \cite{muratov_gnedin10}.}
\vspace{0.3cm}
  \label{fig:gc_met}
\end{figure}

\section{Metallicity Bimodality}
  \label{sec:met}

Figure~\ref{fig:gc_met} shows the metallicity bimodality in the model
of \cite{muratov_gnedin10}.  Note that the model imposes the same
formation criteria for all clusters, without explicitly
differentiating between the two modes.  The only variables are the
gradually changing amount of cold gas, the growth of protogalactic
disks, and the rate of merging.  Yet, the model produces two peaks of
the metallicity distribution, centered at $\feh \approx -1.6$ and
$\feh \approx -0.6$, matching the Galactic globular clusters.

The red peak is not as pronounced as in the observations but is still
significant.  Early mergers of low-mass progenitors contribute only
blue clusters.  Interestingly, later major mergers contribute both to
the red and blue modes, in about equal proportions.  They are expected
to produce a higher fraction of red clusters in galaxies with more
active merger history, such as in massive ellipticals.

In this scenario, bimodality results from the history of galaxy
assembly (rate of mergers) and the amount of cold gas in protogalactic
disks.  Early mergers are frequent but involve relatively low-mass
protogalaxies, which produce preferentially blue clusters.  Late
mergers are infrequent but typically involve more massive galaxies.
As the number of clusters formed in each merger increases with the
progenitor mass, just a few late super-massive mergers can produce a
significant number of red clusters.  The concurrent growth of the
average metallicity of galaxies between the late mergers leads to an
apparent ``gap'' between the red and blue clusters.

Our prescription links cluster metallicity to the average galaxy
metallicity in a one-to-one relation, albeit with random scatter.
Since the average galaxy metallicity grows monotonically with time,
the cluster metallicity also grows with time.  The model thus encodes
an age-metallicity relation, in the sense that metal-rich clusters are
younger than their metal-poor counterparts by several Gyr.  However,
clusters of the same age may differ in metallicity by as much as a
factor of 10, as they formed in the progenitors of different mass.

\begin{figure}[t]
\centering
 \includegraphics[width=2.63in]{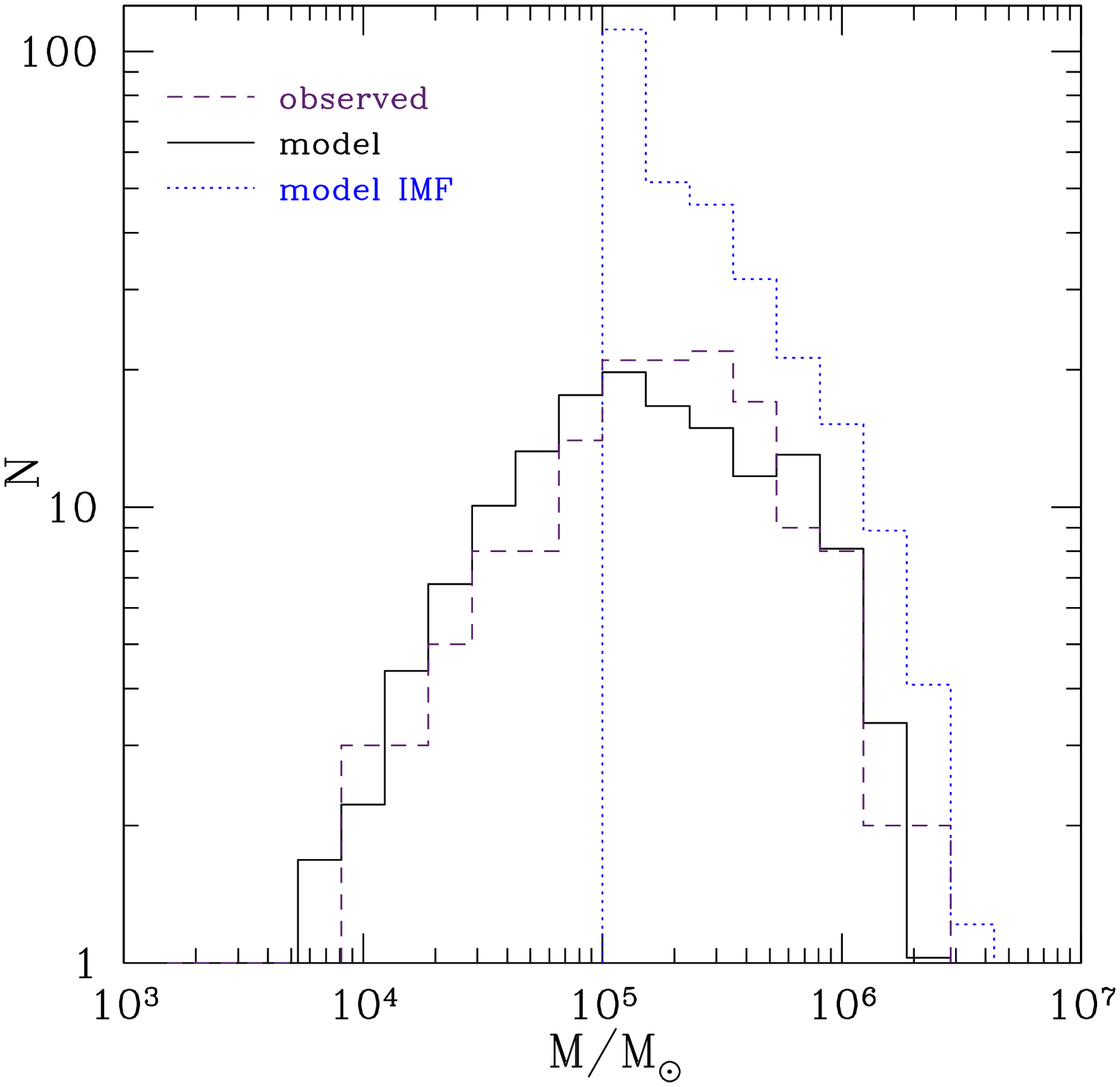}
 \includegraphics[width=2.63in]{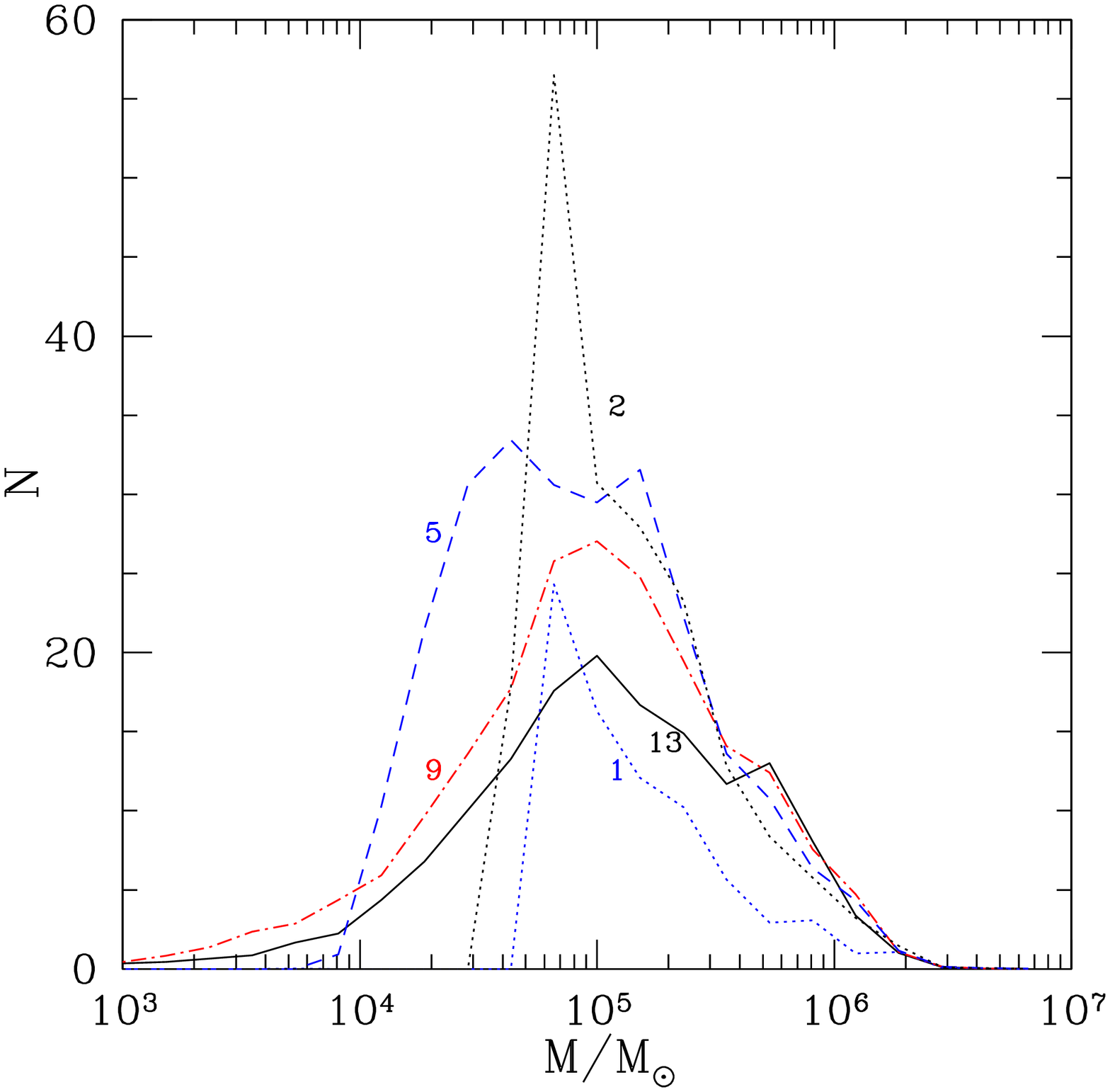}
\vspace{-0.3cm}
\caption{\small {\it Left:} Dynamically evolved model clusters at
  $z=0$ ({\it solid}), compared to the Galactic globular clusters
  ({\it dashed}).  Dotted histogram shows the combined initial masses
  of model clusters with $M > 10^5\, \Msun$ formed at all epochs,
  including those that did not survive until the present. {\it Right:}
  Evolution of the mass function at cosmic times of 1 Gyr ($z \approx
  5.7$, {\it dotted}), 2 Gyr ($z \approx 3.2$, {\it dotted}), 5 Gyr
  ($z \approx 1.3$, {\it dashed}), 9 Gyr ($z \approx 0.5$, {\it
  dot-dashed}), and 13.5~Gyr ($z=0$, {\it solid}).}
\vspace{0.3cm}
  \label{fig:mf}
\end{figure}

\section{Evolution of the Mass Function}

Some of the old and low-mass clusters will be disrupted by the gradual
escape of stars and will not appear in the observed sample.
Figure~\ref{fig:mf} shows the dynamical evolution of the cluster mass
function, as a result of stellar mass loss, tidal truncation, and
two-body evaporation.  Even though the model parameters were tuned to
reproduce the metallicity, not the mass distribution, the mass
function at $z=0$ is consistent with the observed in the Galaxy.
Majority of the disrupted clusters were blue clusters that formed in
early low-mass progenitors.

Right panel of Figure~\ref{fig:mf} illustrates the interplay between
the continuous buildup of massive clusters ($M > 10^5\, \Msun$) and
the dynamical erosion of the low-mass clusters ($M < 10^5\, \Msun$).
Expecting that most clusters below $10^5\, \Msun$ would eventually be
disrupted, we did not track their formation in the model.  Instead,
the low end of the mass function is built by the gradual evaporation
of more massive clusters.  Note that most of the clusters were not
formed until the universe was 2 Gyr old, corresponding to $z \approx
3$.  The fraction of clusters formed before $z \approx 6$, when cosmic
hydrogen was reionized, is small.

An exciting prediction of the model is a high fraction of galaxy
stellar mass locked in star clusters at $z>3$: $M_{GC}/M_* \approx
10-20\%$.  This fraction declines steadily with time and reaches
0.1\% at the present epoch.


\end{document}